# Magnetic Properties of Ternary Gallides of type $R$Ni$_4$Ga ($R$ = Rare earths)

**Devang A. Joshi and C.V. Tomy**
*Department of Physics, Indian Institute of Technology, Mumbai 400076, India*
**D. S. Rana, R. Nagarajan and S. K. Malik**
*Tata Institute of Fundamental Research, Mumbai 400005, India*

The magnetic properties of $R$Ni$_4$Ga ($R$ = La, Pr, Nd, Sm, Gd, Tb, Dy, Ho, Er, Tm and Lu) compounds have been investigated. These compounds form in a hexagonal CaCu$_5$ type structure with a space group $P6/mmm$. Compounds with the magnetic rare earths, $R$ = Nd, Sm, Gd, Tb, Dy, Ho, Er and Tm, undergo a ferromagnetic transition at 5 K, 17 K, 20 K, 19 K, 12 K, 3.5 K, 8 K and 6.5 K, respectively. The transition temperatures are smaller compared to their respective parent compounds $R$Ni$_5$. PrNi$_4$Ga is paramagnetic down to 2 K. LaNi$_4$Ga and LuNi$_4$Ga are Pauli paramagnets. All the compounds show thermomagnetic irreversibility in the magnetically ordered state except GdNi$_4$Ga.
**Keywords :** Ferromagnetism & Crystal Field

## Introduction:

During the past few years, the intermetallic compounds with general formula $RT_5$ ($R$ = rare earth and $T$ = transition element) and their substitutional derivatives have been extensively studied because of a variety of interesting properties they exhibit, such as high coercivity, crystal field effects, magnetocaloric effect, spin fluctuation, magnetic anisotropy, etc [1-4]. Most of these compounds crystallize in a hexagonal structure. The existence of a local moment on the transition metal ions in an alloy depend upon the $3d$ orbital overlap, or in other words, the number and type of the nearest neighbors [5]. The magnetic rare earths do not appreciably lose their moment because the $4f$ orbital lies deep within the atom and is less affected by the overlap. This series of compounds provide a wide opportunity to study the interplay between $4f$ and $3d$ sub lattices when both of them are magnetically ordered or either of them is magnetically ordered. In the $R$Ni$_5$ series of compounds, the interplay between the spin orbit coupling, exchange interaction (mainly due to the $R$ ion) and crystalline electric field results in a strong magneto-crystalline anisotropy which prevails at low temperatures. The electronic band structure calculations on the structurally related YCo$_5$ compound [6] show that quite a large change of the magneto-crystalline anisotropy is expected for comparatively small shift of the Fermi energy. A large value of magnetocaloric effect found in these compounds has also attracted a great deal of research, because of their practical application in magnetic refrigeration. The efficiency of a magnetic refrigerator depends on the magnitude of the magnetocaloric effect of the magnetic material used.

We have synthesized and studied one of the substitutional derivative of the $R$T$_5$ series namely $R$Ni$_4$Ga ($R$ = rare earth). In this paper we present a detailed report on the magnetic properties of $R$Ni$_4$Ga ($R$ = La, Pr, Nd, Sm, Gd, Tb, Dy, Ho, Er, Tm and Lu) compounds. All the compounds with magnetic rare earth undergo a ferromagnetic transition at low temperatures except PrNi$_4$Ga. PrNi$_4$Ga is found to be paramagnetic down to 2 K.

Experimental:

Polycrystalline samples of the series $R$Ni$_4$Ga ($R$ = La, Pr, Nd, Sm, Gd, Tb, Dy, Ho, Er, Tm and Lu) were prepared by repeated arc melting of the stoichiometric amounts of the constituent elements on a water cooled cooper hearth in a purified argon atmosphere. Titanium button was used as an oxygen getter. The starting materials were rare earth elements of 99.99%, Ni 99.9% and Ga 99.99% purity. A small amount of extra Ga was added to compensate for the loss during melting. The total weight loss during the arc melting was less then 0.5% and therefore the alloy composition were assumed to remain unchanged from the original stoichiometric ratios. The as-melted buttons were wrapped in Tantalum foil, sealed in an evacuated quartz tube and annealed at 800º C for 15 days. The compound YbNi$_4$Ga could not be formed by us in single phase by the method described. Room temperature powder X-ray diffraction pattern of the samples were obtained using Panalytical X-ray diffractometer equipped with Cu-Kα radiation. Magnetic susceptibility measurements were carried out using SQUID and PPMS magnetometers (Quantum Design, USA). Magnetization *vs* field isotherms were obtained at various temperatures using a VSM (Oxford Instruments, UK).

Results & Discussion:

From a comparison of the x-ray diffraction pattern of the samples before and after annealing, it was found that the intensity of the peaks increases considerably after annealing, even though the x-ray pattern in each case remains the same. This indicates that annealing improves only the homogeneity of the sample. The diffraction patterns of the annealed samples were analyzed using the Rietveld refinement method (FullProf program). All the $R$Ni$_4$Ga compounds are found to crystallize in a CaCu$_5$ type hexagonal structure, with a space group of *P6/mmm* (No.191). The Reitveld refinement for one of the compound, SmNi$_4$Ga is shown in Fig. 1. In the parent $R$Ni$_5$ compounds, Ni can occupy two different crystallographic sites 2$c$ and 3$g$. However, in the Ga substituted compounds, the refinement of the x-ray pattern gives the best agreement with the experimental data when Ga is considered occupying only the 3$g$ site. It is quite possible that Ga substitutes only at the 3g site situated in the $z = ½$ plane in order to maintain the structure, because of the absence of '$R$ ion' in this plane which allows a larger Ni-Ga distance compared to the 2$c$ site [7]. Thus the atomic positions and the occupancies for $R$Ni$_4$Ga compounds obtained from the refinement are: $R$\{1$a$, (0,0,0), Occu.(1)\}, Ni \{2$c$, (1/3,2/3,0), Occu.(1)\}, Ni and Ga (3$g$, (1/2,0,1/2), Occu.(Ni:0.67 and Ga:0.33)\}. The lattice parameters of $R$Ni$_4$Ga compounds are listed in table 1 along with the corresponding parameters for $R$Ni$_5$ compounds for comparison. These lattice parameters of $R$Ni$_4$Ga compounds are greater then the corresponding parent compound RNi$_5$. The reason for the increase in the lattice parameter can be attributed to the large metallic radii of Ga (0.141 nm) compared to Ni (0.124 nm). The unit cell volumes of $R$Ni$_4$Ga and $R$Ni$_5$ compounds are shown in Fig. 2, which clearly indicate the same trend of lanthanide contraction as expected for the trivalent rare earth ions, except for Ce because of its mixed valent behavior in both the series [8].

The temperature dependence of the magnetic susceptibility of $R$Ni$_4$Ga ($R$ = Pr, Nd, Sm, Gd, Tb, Dy, Ho, Er and Tm) compounds in the field cooled (FC) & zero field cooled (ZFC) mode is shown in Fig. 3 & 4. Compounds with $R$ = Nd, Sm, Gd, Tb, Dy, Ho, Er and Tm undergo a ferromagnetic transition at approximately 5 K, 17 K, 20 K, 19 K, 12 K, 3.5 K, 8K and 6.5 K, respectively. The transition temperature is defined by a sharp peak in the ac susceptibility curve (not shown). The paramagnetic susceptibility above the transition temperature for all the magnetic rare earth compounds (except SmNi$_4$Ga) was fitted to the Curie-Weiss law. The paramagnetic Curie temperatures and the effective magnetic moments thus obtained are given in table 1. In case of SmNi4Ga the first excited state of the Sm$^{3+}$ ion lies very near to the ground state ($\Delta E_{7/2-5/2} \approx 1400$ K) hence the susceptibility in the paramagnetic state can not be fitted to the Curie-Weiss Law. The modified Curie-Weiss Law to fit the susceptibility in such cases is given by

$$\chi = \frac{N_A}{k_B}\left(\frac{\mu_{eff}^2}{3(T-\theta_p)} + \frac{\mu_B^2}{\delta}\right)$$

Here $N_A$ is the Avogadro number, $k_B$ is the Boltzmann's constant, $\mu_B$ is the Bohr magneton, $\mu_{eff}$ is the effective magnetic moment in unit of $\mu_B$, $\theta_P$ is the Curie-Weiss temperature and $\delta = 7\Delta E/20$ in which $\Delta E$ is difference between the ground state and the first excited state. For the $Sm^{3+}$ ion the first term in the above equation represent a Curie-Weiss contribution from J = 5/2 ground state, while the second term is the temperature independent Van Vleck correction arising from the accessible first excited J = 7/2 state. In the absence of crystal fields the probable values of $\mu_{eff}$ and $\delta$ are 0.845 $\mu_B$ and 490 K respectively. The values obtained from the susceptibility fit of SmNi$_4$Ga are $\mu_{eff}$ = 0.81 $\mu_B$ and $\delta$ = 270 K. The appreciably low value of $\delta$ may be attributed to the crystal field effects. The effective magnetic moments are close to and less than the theoretically expected values for free $R^{3+}$ ions except CeNi$_4$Ga [8] and TmNi$_4$Ga. The magnetic behavior of LaNi$_4$Ga (Fig. 4*f*) and LuNi$_4$Ga is a typical of a Pauli-Paramagnet. The upturn of the magnetic susceptibility of LaNi$_4$Ga (Fig. 4*f*) at low temperatures may be due to a small amount of impurity phase in our sample. The above results confirm that Ni atoms have no contribution to the effective magnetic moments in this series of compounds. So any contribution from the 3d electrons, if present (may be due to the moment induced by the magnetic rare earth), can be neglected.

According to the well known de-Gennes scaling, the $T_C$ of the isostructural members of the rare earth series is proportional to $(g_J - 1)^2 J(J+1)$, where $g_J$ is the Lande factor and J is the total angular momentum. In the title compounds this behaviour is not closely followed, especially in the case of TbNi$_4$Ga and SmNi$_4$Ga where the $T_C$ deviates appreciably from their expected values (inset of Fig. 2). Such a behavior is found only in SmNi$_5$ in the case of the parent compounds. This is attributed to the crystalline field effects [9]. It has been shown that crystal field of a suitable sign can enhance the ordering temperature of a compound compared to that expected by the de-Gennes scaling [10, 11]. Sometimes it is even possible that the $T_C$ becomes larger than that of the corresponding Gd compound as in the case of TbPdSn [12], SmNi4B [13], etc. The other possible reason especially for the Sm compounds is the strong exchange coupling between the 4*f* electrons and the conduction electrons caused by the hybridization of the 4*f* state with the conduction electron state [14].

In the magnetically ordered state, the $R$Ni$_4$Ga compounds, except GdNi$_4$Ga, exhibit a thermo-magnetic irreversibility. That is, the ZFC and FC magnetization curves follow different paths below $T_C$. The magnetization measured in the FC state at 500 Oe field increases with decrease in temperature, and after a sharp jump at $T_C$ the susceptibility tends to saturate. In case of ZFC state the magnetization with decrease in the temperature follows the same path as that of the FC up to the transition temperature, just after that it deviates from the FC curve followed by maxima and then decreases some what similar to that of an antiferromagnetic type of transition. Since the above behaviour is absent in the Gd compound (the total orbital angular momentum for Gd is zero, implying a minimum magnetic anisotropy), we can effectively say that the thermomagnetic irreversibility in the rest of the compounds arises due to the anisotropy of the rare earth ion. Similar behavior is also seen in the compounds Pr$_2$Fe$_{17}$ [15], Sm$_2$Co$_{17}$ [16] and $R$Ni$_4$Cu [17]. Thermomagnetic irreversibility is well known in Spin-Glass and magnetically frustrated systems. However, such a behavior is also observed in anisotropic ferromagnets. Thermo-magnetic irreversibility in an anisotropic ferromagnetic system can occur, if the alignment of the domains is restricted by an energy barrier. So cooling the specimen without the magnetic field will lead to a pinning of the domain walls. Then subsequent application of the magnetic field results in alignment of some of the domains and increase in the temperature helps in overcoming the energy barrier. Hence the magnetization increases initially with temperature up to a temperature near $T_C$ at which a peak occurs. The peak is a result of a competitive process between field and thermal energy after which the thermal energy dominates and magnetization decreases.

The magnetization isotherms for $R$Ni$_4$Ga ($R$ = Pr, Nd, Sm, Gd, Tb, Dy, Ho, Er and Tm) compounds at 2 K and up to 9 T are shown in Fig. 5. The behavior is similar to that expected for a ferromagnetic material except for PrNi$_4$Ga. The maximum magnetization ($\mu_B$/per formula unit) achieved for $R$ = Pr, Nd, Sm, Tb, Dy, Ho, Er and Tm at 2 K and 9 T are 1.0, 1.76, 0.27, 6.9, 6.2, 6.7, 6.5 and 2.6, respectively. These values are much smaller than the theoretical saturation moment expected for the corresponding free $R^{3+}$ ions. The remnant magnetization is negligibly small in all this compounds except SmNi$_4$Ga and TbNi$_4$Ga. The magnetic isotherms (hysteresis curve) for SmNi$_4$Ga and TbNi$_4$Ga (inset) at 2 K are shown in Fig. 6. TbNi$_4$Ga shows a remnant magnetization of ≈ 4$\mu_B$ and a coercive field of 0.22 T. SmNi$_4$Ga shows a magnetic hysteresis with a coercive field of 4.1 T and a magnetization of 0.27$\mu_B$ at a field of 10 T. This coercive field is obtained for a material annealed for 15 days. Since coercivity is the microstructural property, it depends upon defects, strains, non-magnetic atoms, etc. in the material. It decreases with the annealing of the material. The occurance of such a high coercivity can be explained in terms of narrow domain walls [19, 20]. Narrow domain walls are expected when the anisotropy energy (K) becomes of the order of magnitude or greater then the ferromagnetic coupling energy (C), (K/C > 2/3). Such condition arises in case of the Sm compounds because of the huge magneto-crystalline anisotropy of the Sm$^{3+}$ ion. In case of SmNi$_5$, a very large magnetic anisotropy has been observed in single crystal [21] and the domain wall width has been argued to be as narrow as one atomic layer [22]. In view of this, SmNi$_4$Ga may also have narrow domain walls of few atomic layers. For GdNi$_4$Ga, the saturation magnetization obtained at 2 K and 9 T is 7.25 $\mu_B$/f.u. which is higher than that expected for the free Gd$^{3+}$ ion (7.0 $\mu_B$/f.u.). This supports our explanation of minimum anisotropy in GdNi$_4$Ga and an anisotropic behavior in the rest of the compounds. The saturation moment obtained for GdNi$_5$ is 6.8 $\mu_B$/f.u. and a structurally related GdNi$_4$B is 7.4 $\mu_B$/f.u. The saturation moment in GdNi$_5$ is explained using three different contributions [15]. The main contribution comes from the saturation moment of Gd$^{3+}$ ion (+7 $\mu_B$/f.u.). The second contribution is from the polarization of the incompletely filled 3$d$ band which couples antiparallel to the 4$f$ moment (−0.76 $\mu_B$/f.u.). The third is the positive coupling of the 4$f$ moment with the polarization of the itinerant electrons ((0.56 $\mu_B$/f.u.). In the case of GdNi$_4$B, the negative contribution from the polarization of 3$d$ band is absent because of its possible filling up. This implies that in GdNi$_4$Ga also, the increase in the saturation moment can be attributed to the absence of the 3$d$ contribution. Thus the itinerant-electron polarization contribution in GdNi$_4$Ga (0.25 $\mu_B$/f.u.) can be considered to be less than that in GdNi$_5$ (0.56 $\mu_B$/f.u.). This may be due to the increase in the unit cell volume of GdNi$_4$Ga which in turn decreases the absolute value of the conduction electron polarization and also the $T_C$.

Conclusion:

We have synthesized and studied the magnetic properties of the ternary gallides $R$Ni$_4$Ga ($R$ = rare earths). All the magnetic rare earth compounds show ferromagnetic ordering at low temperatures except PrNi$_4$Ga which follows the Curie-Weiss behaviour down to 2 K. The transition temperatures are smaller compared to their respective parent compounds $R$Ni$_5$. The $T_C$ values of $R$Ni$_4$Ga compounds do not follow the *de-Gennes* scaling. In particular the $T_C$ of SmNi$_4$Ga & TbNi$_4$Ga very appreciably from the expected values and the possible reason is attributed to the crystal field effects. These compounds also show magnetic anisotropy except GdNi$_4$Ga.

**Figure Captions:**

**Fig.1 :** The observed and calculated diffraction pattern of SmNi$_4$Ga along with the difference. The vertical bars indicate Braggs position.

**Fig. 2 :** Comparison of unit cell volume of RNi$_4$Ga & RNi$_5$ series of compounds. The inset shows the comparison of the transition temperature with that expected from the D-Gennes scalling.

**Fig. 3 :** Susceptibility *vs* Temperature plot for RNi$_4$Ga ( R = Tb, Ho, Er, Dy, Nd and Tm) under *FC* ( Filled data points) and *ZFC* (unfilled data points) conditions.

**Fig. 4 :** Susceptibility *vs* Temperature plot for RNi$_4$Ga ( R = Sm, Gd, Pr and La) with SmNi$_4$Ga and GdNi$_4$Ga under *FC* ( Filled data points) and *ZFC* (unfilled data points) conditions. In Fig. *f* right and left scale denotes susceptibility for LaNi$_4$Ga and PrNi$_4$Ga respectively.

**Fig. 5 :** Magnetization *vs* Field plot for RNi$_4$Ga ( R = Pr, Sm, Nd, Gd, Tb, Ho, Er, Dy and Tm) from 0 to 9 T at 2 K. The right scale denotes the magnetization for SmNi$_4$Ga.

**Table. 1**

| R | $R$Ni$_5$ | | $R$Ni$_4$Ga | | | | | |
|---|---|---|---|---|---|---|---|---|
| | $a$(Å) | $c$(Å) | $a$(Å) | $c$(Å) | $\mu_{eff}$ ($\mu_B$) | $\theta_P$(K) | $T_C$(K) | Ref. |
| La | 5.014 | 3.983 | 5.071 | 4.064 | - | - | P-P | * |
| Ce | 4.878 | 4.06 | 4.945 | 4.075 | 0.8 | -35 | P | 8 |
| Pr | 4.957 | 3.976 | 5.016 | 4.057 | 3.57 | -9 | P | * |
| Nd | 4.952 | 3.975 | 5.001 | 4.051 | 3.6 | 0 | 5 | * |
| Sm | 4.924 | 3.974 | 4.965 | 4.046 | 0.81 | 15 | 17 | * |
| Gd | 4.906 | 3.968 | 4.961 | 4.038 | 7.9 | 20.5 | 20 | * |
| Tb | 4.894 | 3.966 | 4.948 | 4.043 | 9.7 | 19 | 19 | * |
| Dy | 4.872 | 3.968 | 4.932 | 4.036 | 10.6 | 11 | 12 | * |
| Ho | 4.872 | 3.966 | 4.929 | 4.041 | 10.6 | 2.3 | 3.5 | * |
| Er | 4.858 | 3.965 | 4.910 | 4.036 | 9.6 | 9.5 | 8 | * |
| Tm | 4.852 | 3.963 | 4.904 | 4.035 | 7 | 5 | 6.5 | * |
| Lu | 4.832 | 3.961 | 4.884 | 4.031 | - | - | P-P | * |

P-P : Pauli Paramagnet,   * : This work,   P : Paramagnet

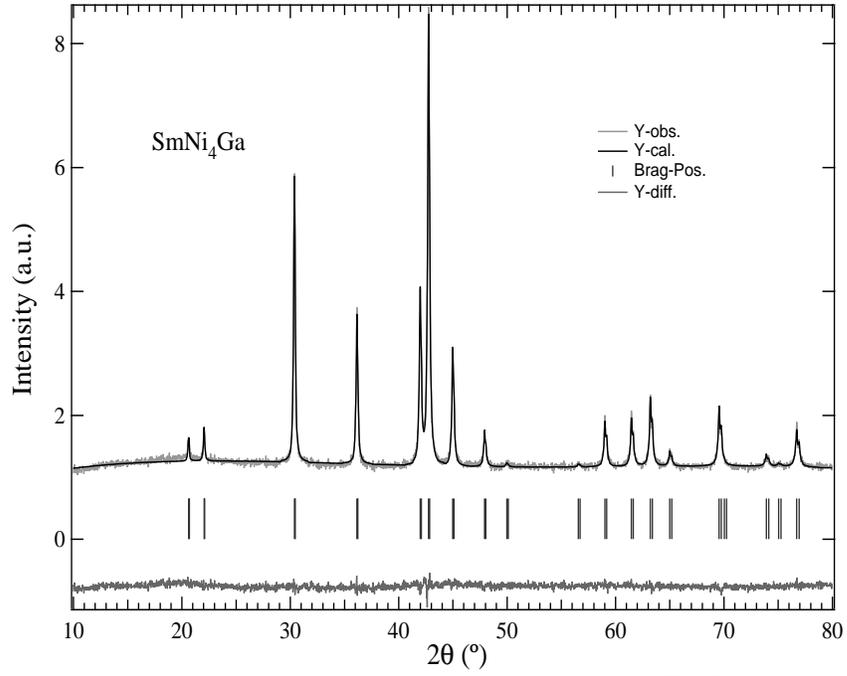

Fig.1 Devang *et al*

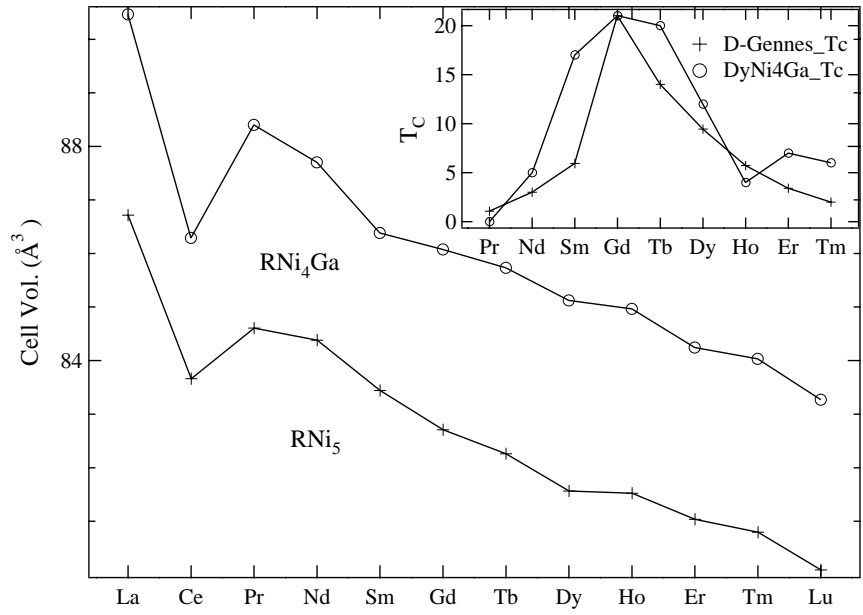

Fig.2 Devang *et al*

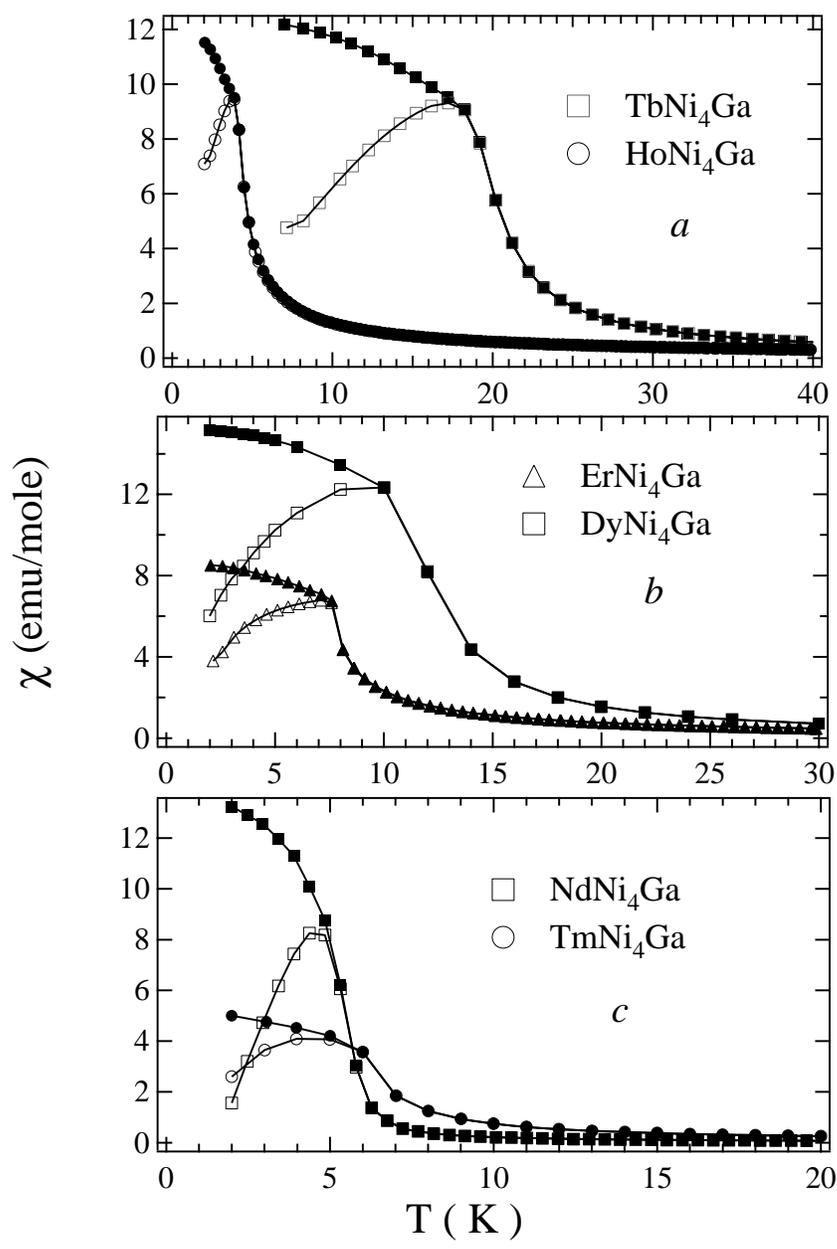

Fig.3 Devang *et al*

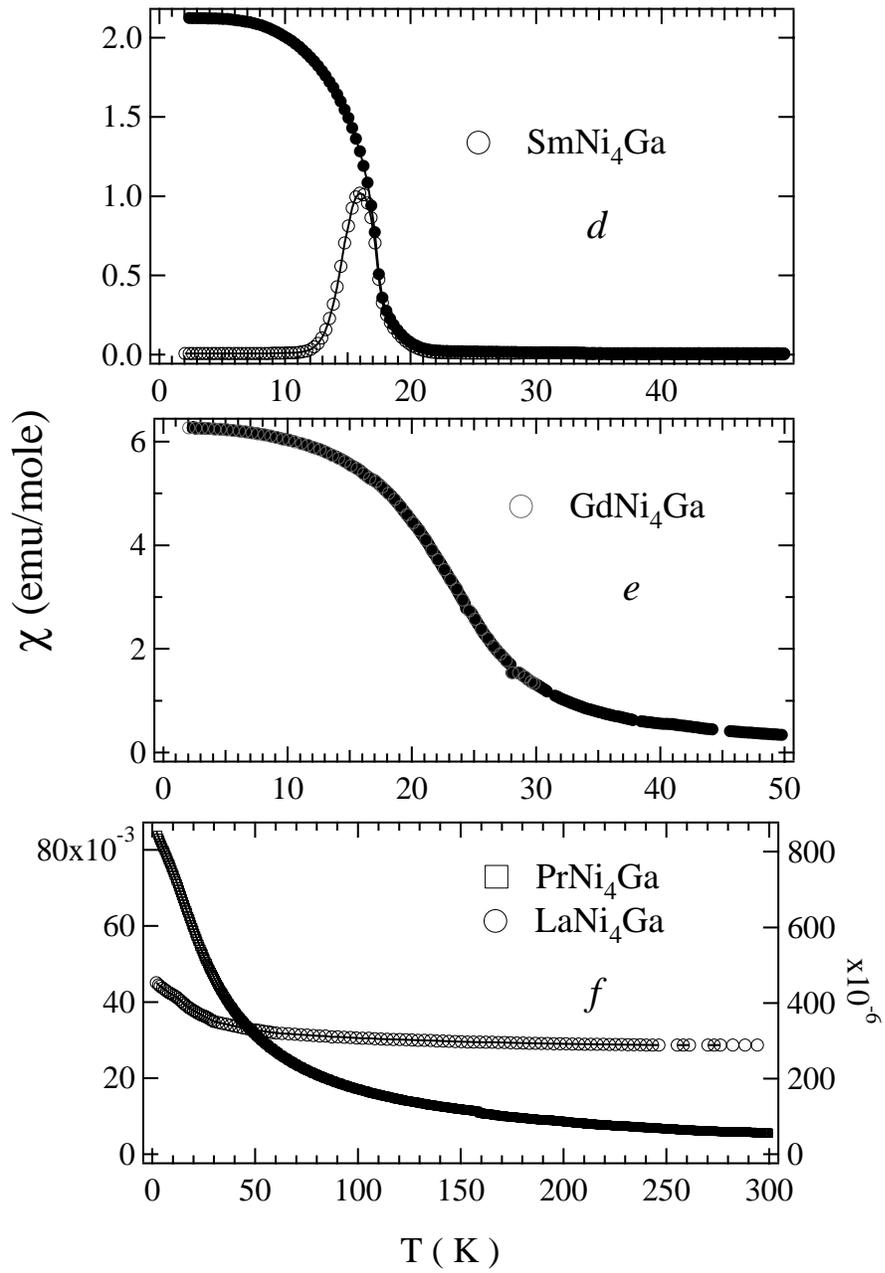

Fig.4 Devang *et al*

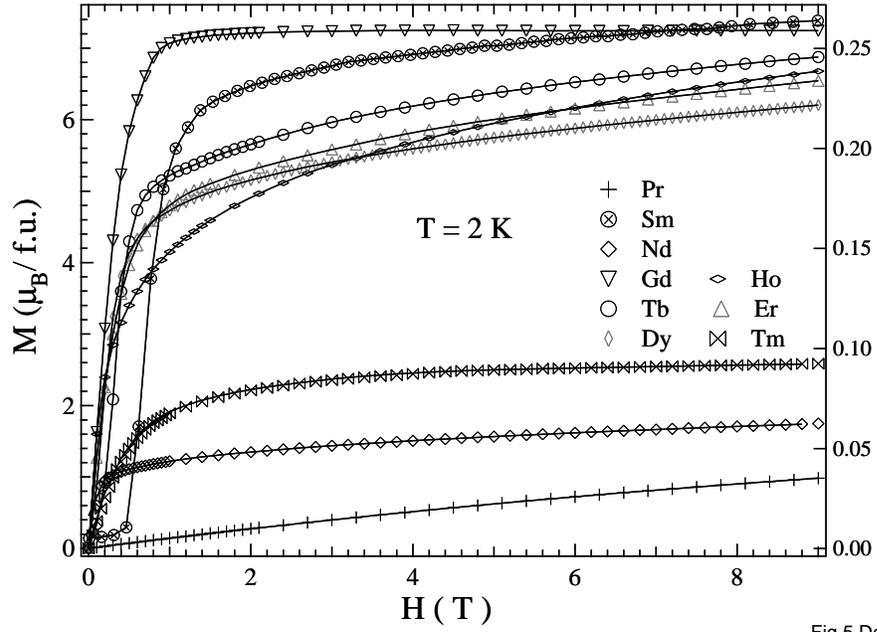

Fig.5 Devang *et al*

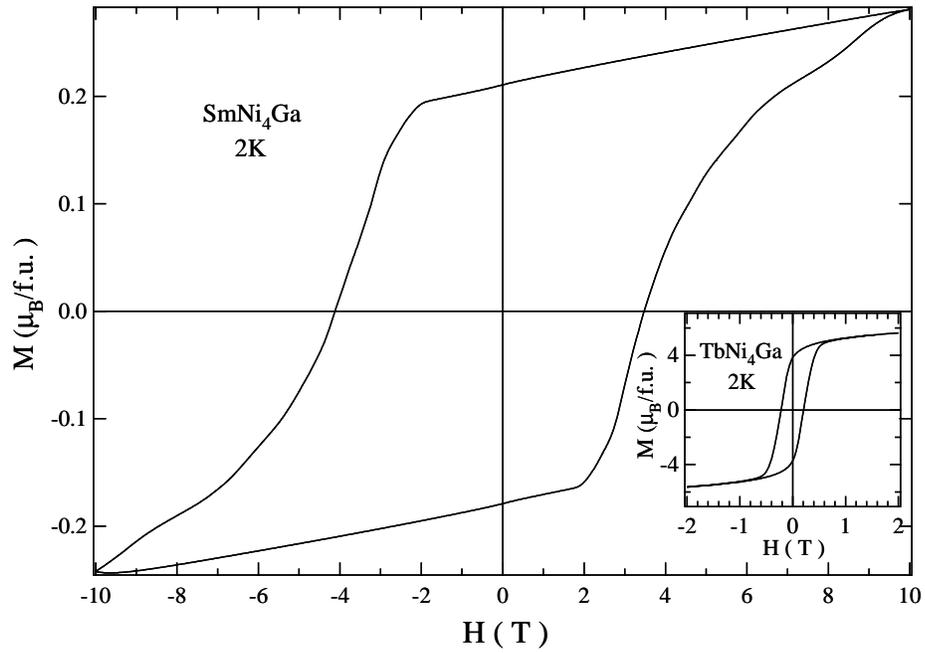